\documentclass[preprint2]{aastex}

\slugcomment{Not to appear in Nonlearned J., 45.}

\shorttitle{Simple Model of the Outflow from Starburst Galaxies}
\shortauthors{V.N. Zirakashvili and H.J.V\"olk}

\begin{document}


\title{Simple Model of the Outflow from Starburst Galaxies: Application to 
Radio Observations}


\author{V.N.Zirakashvili}
\affil{Institute for Terrestrial Magnetism, Ionosphere and Radiowave
Propagation, 142190, Troitsk, Moscow Region, Russia}
\affil{Max-Planck-Institut f\"{u}r\ Kernphysik, Postfach 103980, 69029 
Heidelberg, Germany}
\email{zirak@mpimail.mpi-hd.mpg.de}

\and

\author{H.J.V\"olk}
\affil{Max-Planck-Institut f\"{u}r\ Kernphysik, Postfach 103980, 69029 
Heidelberg, Germany}



\begin{abstract}
We present an analytical model for the hydrodynamic outflow from the 
disk of a starburst galaxy. The model is used to  calculate
the cosmic ray propagation and the radio intensity distribution in the  
nuclear starbust region of NGC 253. We find that the cosmic ray energy 
production rate of the central 600 pc of NGC 253 is about $3\cdot 10^{41}$ erg 
s$^{-1}$, that is about 15 percent of the total mechanical supernova power. 
 For this inner region we estimate a terminal outflow velocity of 900 
km s$^{-1}$ and a mass loss rate of $\dot{M}=2\div 4M_\odot$ per year.
\end{abstract}



\keywords{cosmic rays--
acceleration--
galaxies}


\section{Introduction}

Observations of the soft X-ray and radio continua and the $H_\alpha $
line emission support the idea of
galactic winds from starburst galaxies 
(see e.g. Strickland \cite{strickland04} for 
a review). In these galaxies which exibit a very high and spatially localised stellar 
formation rate, powerful stellar winds and supernova explosions 
produce a large amount of very hot gas. This gas expands out of the starburst 
region and forms a supersonic galactic wind flow. 

Also in more quiescent galaxies such as the Milky Way galactic wind flows  can arise. 
They are primarily driven by cosmic rays 
(Ipavich \cite{ipavich75}; Breitschwerdt et al. \cite{breitschwerdt87},
\cite{breitschwerdt91}, \cite{breitschwerdt93}; Zirakashvili et al. \cite{zirakashvili96}). 
In such galaxies the gas heated by a spatially extended supernova 
activity loses its energy largely by radiative loses, whereas the nonthermal 
cosmic ray component, produced in the same context, can only cool 
adiabatically. In addition, the cosmic rays 
produced in the galactic disk can
not freely escape from the Galaxy but rather amplify Alfv\'en waves 
(e.g. Wentzel \cite{wentzel74}) that lead to an 
efficient dynamical coupling of
the thermal gas to the energetic particle component 
(Ptuskin et al. \cite{ptuskin97}). Therefore the  
cosmic ray pressure gradient is the dominant driver of such an extended 
galactic wind flow. 
Due to the different mechanisms these two types of outflow have somewhat different 
characteristics.   

Chevalier \& Clegg (\cite{chevalier85}) introduced an elegant analytical 
hydrodynamic model 
for the galactic wind flow from a starburst galaxy. It assumes 
steady-state spherical symmetry.  
Despite its simplicity it seems qualitatively 
valid and is often used for estimates of the X-ray 
distribution and of the  cosmic ray propagation. 

However, the assumed spherical shape of the starburst region is a disadvantage 
of this model. In many cases the starburst region has an ellipsoidal shape or 
a disk geometry. They are more adequate for the problem considered. An analytical 
approach to this case is not simple because one should  solve at least 
a two-dimensional system of hydrodynamic equations. 
For this reason a number of numerical calculations
for galactic wind flows have been made (e.g. Habe \& Ikeuchi \cite{habe80}; 
Tomisaka \& Ikeuchi \cite{tomisaka88}; Suchkov et al. \cite{suchkov94}; 
Strickland \& Stevens \cite{strickland00}). 
They take also the temporal evolution of the flow into account. 

These studies evidently show that the powerful energy release in the starburst 
region produces forward and backward shocks propagating in the galactic halo. 
They move mainly perpendicular to the galactic disk. After 3-10 million years 
(depending on the surrounding halo gas density) the shocks break through into 
the intergalactic medium. At smaller distances the galactic wind flow becomes 
steady state. It is bounded by almost vertical walls of the shocked wind and halo 
gas (e.g. Suchkov et al. \cite{suchkov94}).

The hydrodynamic equations can be significantly simplified for so-called 
self-similar stationary flows. This means that different streamlines have the 
same shape but different scales. The equations then reduce to ordinary 
differential equations. This approach was first introduced by Bardeen \& Berger 
(\cite{bardeen78}) for galactic winds. It has been successfully used in the 
theory of collimated rotating magnetohydrodynamic flows from accretion disks and 
of active galactic nuclei (e.g. Blandford \& Payne \cite{blandford82}; 
Contopoulos \& Lovelace \cite{contopoulos94}). In spite of rather stringent 
boundary conditions at the base of the flow in the form of a spatial power-law 
gas pressure dependence, it appears to us as a useful tool. {\bf Indeed} many 
nuclear starbursts do show power-law pressure profiles at distances of 100 to 
1000 pc from the nucleus in the directions perpendicular and parallel to the 
major axis (e.g. Heckman et al. \cite{heckman90}).

In this paper we shall use such a self-similar approach for the description of 
the single-fluid hydrodynamic outflow from a starburst galaxy. We have in 
particular found an explicit nontrivial analytic solution for one special (but 
not unreasonable) power-law dependence of the gas pressure at the base of the 
wind.

We apply the solution found to a calculation of the radio emission from the 
central 600 pc of the starburst galaxy NGC253. The required additional transport 
equation for nonthermal relativistic electrons couples the synchrotron emission 
morphology, intensity, and spectrum with the overall outflow dynamics. This 
permits a determination of the flow speed $u$ and of the cosmic ray energy 
production rate $\dot{E}_\mathrm{CR}$ as a measure of the total mechanical energy 
production $\dot{E}_\mathrm{SN}\sim 10\ \dot{E}_\mathrm{CR}$ in supernovae, 
through a comparison with the radio synchrotron observations.  We limit ourselves 
to this central region because a purely advective propagation model for the 
cosmic ray electrons is used. Such a model would not be adequate for the outer 
part of NGC253 since this outer region exhibits a steep radio spectrum (Hummel et 
al. \cite{hummel84}, Carilli et al. \cite{carilli92}) where diffusion dominates 
advection. Our approach may be considered as complementary to other methods 
related with X-ray and optical spectroscopic observations.     

The basic equations are given in the next section. A one-dimensional 
model for a thin starburst region similar to Chevalier \& Clegg's model is described 
in section 3. Section 4 contains the analytic solution which we found. Applications of 
this solution to cosmic ray propagation in the galaxy NGC 253 are given in 
sections 5 and 6. The outflow parameters derived are described in section 7. 
Section 8 contains a discussion of the results obtained together with our  
conclusions.     

\section{Basic equations}

We shall assume that the starburst region lies in the galactic disk. The problem 
considered is mirror symmetric relative to the galactic midplane. The galactic 
wind flow originates in the disk and extends into the galactic halo. In this 
region the steady-state axially symmetric gasdynamic equations can be written in 
spherical coordinates $R$ and $\theta $ (the polar axis coinciding with the axis 
of galactic rotation):

\begin{equation}
\left( u_R\frac \partial {\partial R}+\frac {u_\theta}{R}
\frac \partial {\partial \theta }\right) u_R=-\frac 1{\rho }
\frac {\partial p}{\partial R}
+\frac {u^2_\theta }R
\end{equation}
\begin{equation}
\left( u_R\frac \partial {\partial R}+\frac {u_\theta}{R}
\frac \partial {\partial \theta }\right) u_\theta =-\frac 1{\rho R}
\frac {\partial p}{\partial \theta }
-\frac {u_Ru_\theta }R
\end{equation}
\[
\left( u_R\frac \partial {\partial R}+\frac {u_\theta}{R}
\frac \partial {\partial \theta }\right) \rho =
\]
\begin{equation}
-\rho \left( \frac 1{R^2}\frac {\partial }{\partial R}(R^2u_R)
+\frac 1{R\sin \theta }\frac {\partial }{\partial \theta }
(\sin \theta u_\theta )\right)       
\end{equation}
\[
\left( u_R\frac \partial {\partial R}+\frac {u_\theta}{R}
\frac \partial {\partial \theta }\right) p =
\]
\begin{equation}
-\gamma p\left( \frac 1{R^2}\frac \partial {\partial R}(R^2u_R)
+\frac 1{R\sin \theta }\frac \partial {\partial \theta }
(\sin \theta u_\theta )\right)       
\end{equation}
Here $u_R$ and $u_\theta $ are the radial and latitudinal  
components of the gas velocity, respectively, $\theta $ is the colatitude, 
$\rho $ and $p$ are the gas mass density and pressure, 
respectively, and $\gamma $ is the adiabatic index of the gas.  
The first two equations are the Euler equations of motion in the 
radial and latitude direction respectively. Eq.(3) 
is the continuity equation and Eq. (4) describes the evolution of the  
gas pressure. 

We neglect gravity, galactic rotation and 
 dynamical effects of the average magnetic field since they are insignificant for 
starburst outflows (cf. Chevalier \& Clegg \cite{chevalier85}). Magnetic fields 
and rotation were taken into account in the framework of self-similar solutions 
for winds from accretion disks and active galactic nuclei. Gravitational 
potentials in a simplified form can also be included (cf. Bardeen \& Berger 
\cite{bardeen78}).

We shall seek the solutions of Eqs. (1)-(4) in the following form:

\begin{equation}
u_R(R, \theta )=u_R(\theta )U(R/g(\theta ))
\end{equation}
\begin{equation}
u_\theta (R, \theta )=u_\theta (\theta )U(R/g(\theta ))
\end{equation}
\begin{equation}
\rho (R, \theta )=n(\theta )\rho _0(R/g(\theta ))
\end{equation}
\begin{equation}
p(R, \theta )=n^{\gamma }(\theta )P(R/g(\theta ))
\end{equation}
Here the functions $U(R_0)$, $\rho _0(R_0)$, and $P(R_0)$ describe the spatial  
dependence of gas velocity, mass density and 
pressure at the base of the wind at $\theta =\pi /2$, respectively, 
the function $g(\theta )$ determines the shape of streamlines, whereas the 
functions 
$u_R(\theta )$, 
$u_\theta (\theta )$ and $n(\theta )$ describe the dependence of the radial, 
latitudinal velocity components and 
gas density along a streamline, respectively. We put $g(\pi /2)=1$ and $n(\pi 
/2)=1$. The 
equation for the streamlines  
is given by $R=R_0g(\theta )$. The equation for the function  $g(\theta )$
may be written as  
\begin{equation}
\frac {\partial g}{\partial \theta }g^{-1}=u_R/u_\theta 
\end{equation}

Substitution of Eqs. (5)-(8) into Eqs. (1)-(4) and taking into 
account Eq. (9) we obtain
\begin{equation}
u_\theta
\frac {\partial u_R}{\partial \theta } =-\frac {n^{\gamma -1}}{\rho _0U^2}
P'R_0+u^2_\theta     
\end{equation}

\begin{equation}
u_\theta 
\frac {\partial u_\theta}{\partial \theta } =-\frac {\gamma P}{\rho _0U^2}n^{\gamma -2}
\frac {\partial n}{\partial \theta }+\frac {n^{\gamma -1}}{\rho _0U^2}P'
R_0\frac {\partial g}{\partial \theta }g^{-1}-u_Ru_\theta      
\end{equation}

\begin{equation}
u_\theta
\frac {\partial n}{\partial \theta } =-n\left( 2u_R+\frac {u_\theta }{\tan \theta } 
+\frac {\partial u_\theta }{\partial \theta }\right)  
\end{equation}
Here $'$ denotes the derivative of the function $P$ with respect to its argument $R_0$. 
Eq.(4) is reduced  to an identity. From Eqs. (9) and (12) 
we find that 
\begin{equation}
ng^{2}u_\theta \sin \theta =\mathrm{const}. 
\end{equation}
In order to separate the variables $R_0$ and $\theta $ we have to assume that at 
the wind base $P\propto \rho _0U^2\propto R^\delta _0$,  with a constant 
$\delta <0$. This means that the gas 
pressure at the base is a power-law function of radius and that the ratio of 
dynamical pressure to the thermal pressure (the square of the flow Mach number) 
at the base is constant. Without loss of generality we can put $\rho _0U^2/P=1$. 
Note that $\rho _0(R_0)$ and $U(R_0)$ need not individually follow a power-low in 
$R_0$, but rather only the product $\rho _0U^2$. Now Eqs. (10) and (11) are 
reduced to
\begin{equation}
u_\theta
\frac {\partial u_R}{\partial \theta } =-\delta n^{\gamma -1}+u^2_\theta     
\end{equation}

\[
\left( u^2_\theta -\gamma n^{\gamma -1}\right) u^{-1}_\theta 
\frac {\partial u_\theta}{\partial \theta }  
\]
\begin{equation}
=n^{\gamma -1}\left(
\frac {\gamma }{\tan \theta }
+\frac {u_R}{u_\theta }(\delta +2\gamma )\right) -u_Ru_\theta      
\end{equation}

Using Eqs. (9) and (13)-(15) one integral of the system can be found as 

\begin{equation}
\frac {u^2_R}2+\frac {u^2_\theta }2+\frac \gamma {\gamma -1}n^{\gamma -1}= \mathrm{const}
\end{equation}   

The Bernoulli constant for 
the each  streamline differs from this integral by the factor $U^2(R/g(\theta 
))$. 

As a result the problem is reduced to the solution of two ordinary differential equations 
(14) and (15) with the additional integral (16). If the solution is known, the function 
$g$ can be found from Eq. (13). Boundary conditions should be determined 
at the wind base at $\theta =\pi /2$. Generally speaking, for arbitrary boundary 
values $u_R$ and $u_\theta $, the solutions considered do not occupy the whole 
space 
between $\theta =0$ and $\theta =\pi /2$ (cf. Bardeen \& Berger \cite{bardeen78}). 
The boundary values should be adjusted such that the solution  
occupies the whole space. 

As a natural boundary value we shall use $u_R=0$ at $\theta =\pi /2$ for a starburst 
wind. It corresponds to a thin starburst region in the galactic disk. One can 
expect that the hot gas expands perpendicular to the disk in that case.    
 
\section{Solution inside a thin starburst region} 

If the starburst region has the form of thin disk, the one-dimensional hydrodynamic 
equations are a good approximation: 

\begin{equation}
\frac {\partial }{\partial z}\rho u=q(z)
\end{equation}
\begin{equation}
\rho u\frac {\partial u}{\partial z}=-\frac {\partial p}{\partial z}-q(z)u
\end{equation}
\begin{equation}
\frac {\partial }{\partial z}\rho u
\left( \frac {u^2}{2}+\frac {\gamma }{\gamma -1}\frac p{\rho }\right) =Q(z)
\end{equation}
Here $z$ is the height above disk midplane, $q(z)$ and $Q(z)$ are the sources of 
mass and energy, respectively.  
These equations are similar to the 
ones used by Chevalier \& Clegg (\cite{chevalier85}) but for 
a plane parallel case. We shall assume that the problem is symmetric 
relative to the galactic midplane. 

Eqs. (17)-(19) can be integrated:

\begin{equation}
\rho u=\int ^z_0dz'q(z')
\end{equation} 
\begin{equation}
p=p(0)-u\int ^z_0dz'q(z')
\end{equation}
\begin{equation}
\frac {u^2}{2}+\frac {\gamma }{\gamma -1}\frac p{\rho } 
=\frac {\int ^z_0dz'Q(z')}{\int ^z_0dz'q(z')}
\end{equation}

Here $p(0)$ is the gas pressure in the midplane. 
The substitution of $\rho $ and $p$ from the first two equations into the third gives a 
quadratic equation for the gas velocity $u$. Its solution is given by 

\[
u(z)=2\frac {\gamma -1}{\gamma p(0)}\int ^z_0dz'Q(z')
\]
\begin{equation}
\times \left( 1+
\sqrt{1-2\frac {\gamma ^2-1}{\gamma ^2p^2(0)}\int ^z_0dz'Q(z')\int ^z_0dz'q(z')}
\right) ^{-1}
\end{equation}

The total energy and mass production rate per unit area of starburst disk are 
given by 

\begin{equation}
Q_0=2\int ^h_0dzQ(z), 
\end{equation}
\begin{equation}
q_0=2\int ^h_0dzq(z). 
\end{equation}
Here $h$ is the halfwidth of the starburst disk.  
We can calculate the midplane gas pressure $p(0)$ if the Mach number 
$M=\sqrt{\rho u^2/\gamma p}$ at the boundary of the starburst disk $z=h$ is given. 
Its value is determined by the wind solution in the halo region (cf. next Section). 
E.g. $M=1$ in the Chevalier \& Clegg model (\cite{chevalier85}). Using Eqs. (20)-(22) 
we found the gas pressure, the gas mass density and the gas velocity at the 
boundary of 
the starburst disk

\begin{equation}
p(h)=\frac {p(0)}{1+\gamma M^2}=\frac {1}{2\gamma M}
\sqrt{\frac {2(\gamma -1)Q_0q_0}{2+(\gamma -1)M^2}}
\end{equation}
\begin{equation}
\rho (h)=\frac {1}{2M}
\sqrt{\frac {q^3_0}{Q_0}\frac {2+(\gamma -1)M^2}{2(\gamma -1)}}
\end{equation}
\begin{equation}
u(h)=M
\sqrt{\frac {Q_0}{q_0}\frac {2(\gamma -1)}{2+(\gamma -1)M^2}}
\end{equation}
These  quantities formally coincide with the corresponding  
quantities at the wind base in the thin disk approximation: 
$p(h)=p(\pi /2)$, $\rho (h)=\rho (\pi /2)$ and $u(h)=-u_\theta (\pi /2)$.
We should underline that this solution is rather general and is not related to 
self-similarity of the halo outflow. It is valid for any distribution of energy and mass 
sources $Q_0(R_0)$ and 
$q_0(R_0)$ in the galactic disk. However, the value of $M$ in general depends on 
$R_0$ and is determined by a solution in the halo region. For a starburst outflow 
$M$ is 
close to 
unity if the adiabatic index is not close to unity (see also the next Section).  

\section{Analytical solution in the galactic halo} 

In the case $\delta =-2\gamma /(2\gamma -1)$ an analytical solution of Eqs. (14)-(16) 
can be found in the first quadrant $0<\theta <\pi /2$: 

\begin{equation}
u_R=u_\infty \cos \theta 
\end{equation}
\begin{equation}
u_\theta =-(1-\gamma ^{-1})u_\infty \sin \theta  
\end{equation}
\begin{equation}
n=\sin ^{2/(\gamma -1)}\theta 
\end{equation}
\begin{equation}
g=\sin ^{-\gamma /(\gamma -1)}\theta 
\end{equation}
Here the value of $u_\infty $ fixes the asymptotic velocity at distances $R$ 
that are large compared to the radius $R_0$ for each streamline $u_\infty 
U(R/g(\theta ))$. The flow lines for this solution for $\gamma =5/3$ and $\delta 
=-10/7$ and the surfaces of constant Mach number are shown in Fig.1 in 
cylindrical coordinates $r$ and $z$. The initial Mach number of the flow is 
$M=\sqrt{2(\gamma -1)/(2\gamma -1)}$. For other values of $\delta $ the solutions 
of Eqs. (14)-(16) can be obtained numerically. The streamlines for these 
solutions look similar to lines shown in Fig.1. For $\delta $ closer to zero the 
streamlines are closer to the axis, i.e. the wind is collimated more strongly. We 
have not found the solutions for $\delta <-2\gamma /(2\gamma -1)$ and for a 
natural boundary condition $u_R=0$ at $\theta =\pi /2$. Therefore the analytical 
solution with $\delta =-2\gamma /(2\gamma -1)$ has the maximally opened flowlines 
in the framework of self-similar flows.
\begin{figure}[t]
\includegraphics[width=7.5cm]{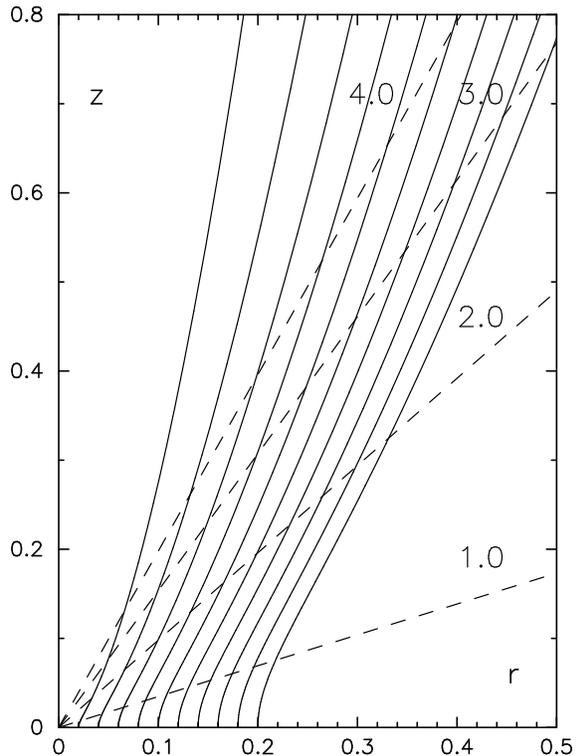}
\caption{Flow lines of the analytical solution ({\it solid lines}) 
for $\gamma =5/3$ and $\delta =-10/7$ 
and surfaces of  constant Mach number ({\it dashed lines}). 
The units of the 
cylindrical coordinates $r$ and $z$
are arbitrary. The Mach number values are shown near the corresponding {\it 
dashed lines}. }
\end{figure}

As was first noted by Bardeen \& Berger (\cite{bardeen78}) 
the equations for self-similar flow do not 
contain any particular points corresponding to sonic points of the flow. The only 
particular point is the point where the sonic velocity equals the latitudinal 
component of 
the flow velocity (cf. Eq. (15)). This component is always smaller than 
the sonic velocity in 
the solutions found. The wind starts at the base with subsonic velocity (initial Mach 
number $M\approx 0.756 $ for $\gamma =5/3$). It is accelerated and passes through 
the sonic point at about $\theta \approx 70^\circ $ (see Fig.1). This picture corresponds 
to a flow in a Laval nozzle. The cross-section of the Laval nozzle shrinks in the 
subsonic part of the flow in order to reach sonic velocities and expands 
after the sonic point in order to obtain an acceleration of the flow beyond. 
As can be seen in Fig.1 this is indeed the case. It is interesting that this 
flow geometry is organized by the gas itself through the perpendicular force balance. 
At the base of the wind the gas pushes the flowlines out of the axis. At larger heights the 
situation is reversed. The wind originating closer to the axis expands faster. The gas 
pressure drops adiabatically and the gas pressure of the wind originating at larger 
distances pushes the flow lines to the axis.  
      
\section{Propagation of cosmic ray electrons in the wind}

We shall neglect the diffusion of energetic electrons in the starburst outflow. 
This seems reasonable since the flow velocity is quite high. In addition, the 
cosmic ray scattering produced by self-excited Alfv\'en waves (Ptuskin et al. 
\cite{ptuskin97}) should be very effective in starbursts. This effect reduces the 
cosmic ray diffusion coefficient. The estimates of this effect in our Galaxy give 
for the diffusion coefficient the value $10^{28}$ cm$^2$ s$^{-1}$ for an energy 
of 10 GeV (cf. Ptuskin et al. \cite{ptuskin97}). It is inversely proportional to 
the cosmic ray production power per unit area of the galactic disk and is 
proportional to the magnetic field strength. Since starbursts have a high cosmic 
ray production power per unit area, typically a factor of 100 to 1000 or more 
larger than in our Galaxy, and since the magnetic field is a factor of 10 larger, 
the diffusion coefficient in starbursts is smaller by at least a factor of 10 to 
100, and the diffusion is negligible in comparison with advection for 10 GeV 
electrons which produce radio synchrotron emission ( for an application of 
these ideas to TeV gamma-ray emitting particles, see Aharonian et al. 
\cite{aharon05}). The cosmic ray diffusion coefficient of much higher energy 
particles can in addition become quite small, if one takes into account 
that the starburst region is very turbulent, 
leading to additional strong magnetic field fluctuations which are 
produced by the turbulent velocity field.

Assuming azimuthal symmetry and a steady state the isotropic part of the cosmic 
ray energy distribution function $N(R,\theta ,E)$ obeys the following equation 

\[
\left( u_R\frac \partial {\partial R}+\frac {u_\theta}{R}
\frac \partial {\partial \theta }\right) N =\frac {\partial }{\partial E}(b(E)N)
\]
\begin{equation}
+ \frac {E^3}3\frac {\partial }{\partial E}\frac N{E^2}
\left( \frac 1{R^2}\frac \partial {\partial R}(R^2u_R)
+\frac 1{R\sin \theta }\frac \partial {\partial \theta }
(\sin \theta u_\theta )\right) 
\end{equation} 
Here the second term on the right hand side describes the adiabatic energy losses, 
while $b(E)$ describes all other energy losses of particles. The boundary condition at the 
wind base $\theta =\pi /2$ can be written as  

\begin{equation}
u_\theta \frac {E^3}3\frac {\partial }{\partial E}\frac {N_0}{E^2}-
\frac \partial {\partial E}(b_d(E)N_0(E))=\frac 12q_{CR}(E)  
\end{equation}
Here $N_0(E)$ is the electron energy distribution at the wind base, $b_d(E)$ describes 
non-adiabatic energy losses in the thin starburst region and  
$q_{CR}(E)$ is the differential source energy spectrum  
of cosmic ray electrons normalized per unit area of the galactic disk.  

The synchrotron emissivity $\epsilon _\nu $ can be written as 

\begin{equation}
\epsilon _\nu =\frac {\sqrt{3}B_\perp e^3}{4\pi mc^2}\int dEN(E)f(\nu /\nu _c)
\end{equation}
Here $e$ and $m$ are the charge and the mass of an electron respectively, $B_\perp $ 
is the strength of magnetic field component perpendicular to the line of sight, $\nu $ is 
the radio-frequency, $c$ is the light velocity and 
$f$ is the function that describes the electron synchrotron emission 
(cf. Landau \& Lifshitz \cite{landau75}). The characteristic frequency of synchrotron 
radiation $\nu _c$ is given by the formula 

\begin{equation}
\nu _c=\frac {3eB_\perp }{4\pi mc}\left( \frac {E}{mc^2}\right) ^2
\end{equation}    

The synchrotron and inverse Compton energy losses can be written as 

\begin{equation}
b(E)=\frac 43c\sigma _T\left( U_{rad}+\frac {B^2}{8\pi}\right)
\left( \frac {E}{mc^2}\right) ^2,
\end{equation}
where $\sigma _T=8\pi e^4/(3m^2c^4)$ is the Thompson cross-section, $B$ is 
the magnetic field strength and $U_{rad}$ is the radiation energy density. 

We shall neglect bremsstrahlung and ionisation losses of electrons in the 
galactic wind flow. This is well justified, since the wind density is small. 
However, these losses may be non-negligible in the galactic disk which contains a 
large mass of gas. They are described by the second term on the left hand side of 
the boundary condition (34). The estimates show that these losses can be 
significant in the very center of a starburst. This depends on the gas mass that 
is rather uncertain. In addition it is not clear, whether radio electrons can 
well penetrate into gas clouds or whether they are quickly transported around 
clouds by the hot rarefied wind. The cosmic ray penetration into the gas clouds 
in the starburst region is beyond the scope of this paper. For the sake of 
simplicity we shall neglect these energy losses in the disk also.

The function $f$ can be expressed in terms of the Mc-Donald function. 
A so-called $\delta $-function approximation is often used for the function $f$. We 
shall use a more accurate method and approximate $f$ as (Pacholczyk \cite{pacholczyk70}): 
\begin{equation}
f(x)=1.81 x^{1/3}\mathrm{e}^{-x},
\end{equation} 
 where $x=\nu /\nu_c$. 
 
The synchrotron intensity observed in a particular direction is given by the integral 
of the emissivity along the line of sight:

\begin{equation}
J_\nu =\int dl \epsilon _\nu 
\end{equation}
 
The spectrum of sources of cosmic ray electrons is given by 

\begin{equation}
q_{CR}(E)=\frac {K^{-1}Q_{CR}(R_0)}{E^2\ln (E_{\max }/E_{\min })}
\end{equation}   
where $Q_{CR}$ is the total power that is transfered by galactic sources to high energy 
cosmic ray protons with energies between $E_{\min }$ and $E_{\max }$ per unit area of the 
galactic disk, and $K$ is the proton to electron ratio in the source. It is expected that 
such a spectrum is formed in supernova remnants by the diffusive shock acceleration mechanism 
(Krymsky \cite{krymsky77}, Axford et al. \cite{axford77}, Bell \cite{bell78}, 
Blandford \& Ostriker \cite{blandford78}). 

We shall neglect the production of secondary electrons and positrons by the 
nucleonic cosmic ray component in the dense gas disk. Although this effect can be 
essential for starbursts (cf. Paglione et al. \cite{paglione96}), its magnitude 
is rather uncertain because of the uncertain gas mass and the uncertain degree of 
penetration of the $10\div 100$ GeV cosmic nucleons into the gas clouds. If 
significant, this effect can be formally included in Eq. (40) by changing the 
proton to electron ratio $K$.

The solution of equation (33) with velocity (5),(6) and energy losses 
$b(R,\theta ,E) =b_0(R,\theta )E^2$ can be written as

\begin{equation}
N(R,\theta ,E)=N_0(E_0)\frac {E^2_0}{E^2}n^{4/3}(\theta )
\end{equation}

Here  $E_0=E_mE/(E_m-E)$ and the maximum energy $E_m $ of electrons is given by 

\[
E^{-1}_m(R,\theta )=-R\int ^{\pi /2}_{\theta }d\theta '\frac {b_0(Rg(\theta ')/g(\theta ), \theta ')}
{U(R/g(\theta ))u_\theta (\theta ')}
\]
\begin{equation}
\times \left( \frac {n(\theta ')}
{n(\theta )}\right) ^{1/3}\frac {g(\theta ')}{g(\theta )}
\end{equation}

Using the analytical solution (30) and (32) and the source spectrum (40) we obtain
\[
N(R,\theta ,E)=\frac {3K^{-1}Q_{CR}(R\sin ^{\gamma /(\gamma -1)}\theta )}
{8E^2\ln (E_{\max }/E_{\min })}
H(E_m-E)
\]   
\begin{equation}
\times \frac {
\sin ^{8/3(\gamma -1)}\theta \gamma /(\gamma -1)}
{u_\infty U(R\sin ^{\gamma /(\gamma -1)}\theta )}
\end{equation}  
where $H(E)$ is the step function. The maximum energy $E_m$ is reduced to 

\[
E^{-1}_m=\frac {R\gamma /(\gamma -1)}{u_\infty U(R\sin ^{\gamma /(\gamma -1)}\theta )}
\]
\begin{equation}
\times \int ^{\pi /2}_{\theta }\frac {d\theta '}{\sin \theta '}
b_0\left( R\left( \frac {\sin \theta }{\sin \theta '}\right) ^{\frac {\gamma }{\gamma -1}}, 
\theta '\right) 
\left( \frac {\sin \theta }{\sin \theta '}\right) ^{\frac {3\gamma -2}{3(\gamma -1)}} 
\end{equation}

\section{Modeling of the radio spectra of NGC 253} 

NGC 253 is a nearby edge-on (inclination angle $i=78^{\circ }$; Pence \cite{pence81}) 
spiral starburst galaxy at distance $d=2.5$ Mpc (Mauersberger et al. \cite{mauersberger96}). 
It has extended radio 
(cf. Carilli et al. \cite{carilli92}) and X-ray 
(cf. e.g. Strickland \cite{strickland04}) halos that 
appear to be related with the powerful superwind flow originating in the 
galactic nucleus. Our model of the outflow can be used to model the spatial 
X-ray and radio distribution in the central region of this galaxy. 

We have performed the modeling of the radio distribution in the central 600 pc of 
NGC 253 using the results obtained in the previous sections. The radio spectrum 
of this galaxy is flat at distances $R_0<1$ kpc from the nucleus (Carilli et al. 
\cite{carilli92}) with a spectral index larger than -0.6. We expect that in this 
region cosmic rays propagate advectively. Concretely, we choose the function 
$U(R_0)=1$. This means that the initial velocity is constant along the base of 
the wind and that the velocity depends on the colatitude $\theta $ only. The 
analytical solution found in the previous sections can be written for this case 
as

\begin{equation}
\rho (R,\theta)=\rho_g\sin ^{-2\frac {\gamma -1}{2\gamma -1}}\theta 
\left( \frac {R}{R_g}\right) ^{-\frac {2\gamma }{2\gamma -1}}
\end{equation}     
\begin{equation}
p(R,\theta )=\frac {1}{2\gamma ^3}(2\gamma -1)(\gamma -1)\rho _gu^2_\infty 
\sin ^{\frac {2\gamma }{2\gamma -1}}\theta \left( \frac {R}{R_g}\right) ^{-\frac {2\gamma}{2\gamma -1}}
\end{equation}
Here the asymptotic velocity $u_\infty $ is the same for all streamlines, $R_g$ is the radius of 
the wind base and $\rho _g$ is the gas density at the wind base at $R=R_g$. Dependence of velocity 
components on colatitude is given by Eqs. (29) and (30). 

We use a radiation field distribution at $R<R_g$ in the simplified form:

\begin{equation}
U_{\mathrm{rad}}=\frac {L_{\mathrm{IR}}+L_\mathrm{B}}{4\pi R^2_gc}
\left( \frac {R}{R_g}\right) ^{-\frac {2\gamma }{2\gamma -1}}
\end{equation}
Here $L_{\mathrm{IR}}$ and $L_{\mathrm{B}}$ are 
the infrared and starlight luminosities of the galaxy inside 
the region with wind base radius. It is reasonable to assume that radiation field is distributed 
similar to distribution of energy production in the starburst disk. 

For a calculation of the synchrotron emissivity we need to define the magnetic 
field distribution. Radio observations of NGC 253 (Beck et al. \cite{beck94}) and 
many other galaxies show that the radio emission is weakly polarized. This means 
that the random component of the magnetic field is larger then the regular 
component. In the case of a starburst this random component can be generated by 
turbulence in the starburst disk and transported by the wind into the halo. 
Another possibility is to treat this random magnetic field as large amplitude 
Alfv\'en waves propagating along the weak regular field and generated by the 
motions of the footpoints of the field lines in the starburst disk. This picture 
is similar to the case of the solar corona.

We shall treat the random magnetic field as a fluid with an adiabatic index $\gamma _m$. It is 
known that Alfv\'en waves have the adiabatic index $\gamma_m=3/2$ 
(Dewar \cite{dewar70}, McKee \& Zweibel \cite{mckee95}); an isotropic 
random magnetic field has the 
adiabatic index $\gamma _m=4/3$ (Mestel \cite{mestel65}).   
We used the following 
expression for the magnetic field which has a  similar R-dependence as 
$p(R,\theta)$ (cf. Eq. (46)):
\begin{equation}
B=B_0(\sin \theta )^{\frac 1{\gamma -1}
\left( \gamma _m-\frac {\gamma ^2}{2\gamma -1}\right) }\left( \frac R{R_g}\right) 
^{-\frac {\gamma }{2\gamma -1}}
\end{equation}    
Here $B_0$ is the magnetic field strength at $R=R_g$. The magnetic field 
component $B_\perp $ in Eqs. (35) and (36) was calculated as $B_\perp 
=\sqrt{2/3}B$. 

The distribution of cosmic ray sources in the starburst disk in Eq. (43) was 
taken in a similar form

\begin{equation}
Q_{\mathrm{CR}}(R_0)=\frac {\gamma -1}{2\gamma -1}\frac {L_{\mathrm{CR}}}{\pi R^2_g}
\left( \frac {R_0}{R_g}\right) ^{-\frac {2\gamma }{2\gamma -1}}, 
\end{equation} 
where $L_{\mathrm{CR}}$ is the cosmic ray energy production rate of the starburst. 

The continuum radio emission of galaxies consists of two main components: 
synchrotron emission of cosmic ray electrons and a thermal emission of the 
ionized gas which is produced mainly in HII regions, ionized by OB stars (see 
e.g. Condon \cite{condon92} for a review). The thermal emission is closely 
related with thermal (so-called free-free) absorption which can be significant at 
low frequencies and can produce a flattening and even a turnover of the spectrum. 
It appears that this indeed the case for the spectrum of the nucleus of NGC 253 
(Carilli \cite{carilli96}). Other possible reasons for the flattening are 
reacceleration of low-energy electrons (Pohl \& Schlickeiser \cite{pohl92}), and 
ionisation and bremsstrahlung energy losses of electrons in the starburst disk 
(Hummel, \cite{hummel91}). We give preference to free-free absorption here, since 
the thermal component of the radio emission is undoubtedly present in NGC 253.

To take 
free-free absorption into account we shall assume that the thin galactic disk seen face-on 
has the following distribution 
of the absorption depth
\begin{equation}
\tau (R_0)=\tau _0 \left( \frac {R_0}{R_g}\right) ^{-\frac {2\gamma }{2\gamma -1}} 
\left( \frac {1\ \mathrm{GHz}}{\nu }\right) ^{2.1}.  
\end{equation}
Here $\tau_0$ is the absorption depth of the disk at 1 GHz and at $R_0=R_g$. It 
was assumed that this distribution is similar to the distribution of sources of 
energy in the galactic disk. This is reasonable, since the absorbing gas is 
ionized by OB stars. The number of these stars is proportional to the input of 
energy. The inclined disk of NGC 253 absorbs the synchrotron emission of the 
northern part of the halo.

The thermal emission brightness of the inclined disk can be expressed in terms of 
the absorption depth $\tau $ and the inclination angle $i$:
\begin{equation}
J_{\mathrm{th}}=2\frac {T\nu ^2}{c^2}\left( 1-\mathrm{e}^{-\tau /|\cos{i}|}\right) . 
\end{equation} 

Here $T$ is the absorbing gas temperature in energetic units. Generally speaking, 
the two last equations are valid if the absorbing gas has a large covering 
factor. This is possibly not the case in the disk of NGC 253. However it is 
possible in the central region, where the main part of the absorption occurs.

We should take into account the synchrotron emission of supernova remnants (SNRs) 
in the starburst disk. The SNRs in the disk have a small volume filling factor. 
However, the magnetic field strength and the number density of relativistic 
electrons in SNRs are larger in comparison with the corresponding quantities in 
the interstellar medium of the starburst disk and in the wind halo. We used the 
following expression for the brightness of the radio emission produced by the 
SNRs from the inclined disk:

\[
J_\mathrm{SNR}=\frac {\gamma -1}{2\gamma -1}
\frac {J_0d^2}{\pi R_g^2\tau }
\left( \frac {R_0}{R_g}\right) ^{-\frac {2\gamma }{2\gamma -1}} 
\left( \frac {\nu }{1\ \mathrm{GHz}}\right) ^{-0.5}
\]
\begin{equation}
\times \left( 1-\mathrm{e}^{-\tau /|\cos{i}|}\right) .
\end{equation} 
This emission from SNRs is also influenced by free-free absorption. The total 
radio 
flux $J_0$ from SNRs is 
about one tenth of the total radio flux for normal galaxies (Condon \cite{condon92}), but 
it can be larger in galaxies with a fast escape of electrons. A more 
quantitative estimate was obtained by Ulvestad (\cite{ulvestad82}) using the relation between 
radio surface brightness and diameter (the $''\Sigma -D''$ relation, see also 
Berezhko \& V\"olk \cite{ber04}): 

\[
J_0=206\ \mathrm{Jy}\ 
\left( \frac d{\mathrm{Mpc}}\right) ^{-2}
\left( \frac {\nu _\mathrm{SN}}{1\ \mathrm{yr}^{-1}}\right) 
\]
\begin{equation}
\times \left( \frac {E_\mathrm{SN}}{10^{50}\ \mathrm{erg}}\right) ^{-1/17} 
\left( \frac {n_\mathrm{ISM}}{1 \ \mathrm{cm} ^{-3}}\right) ^{-2/17} 
\end{equation}  
where $\nu _\mathrm{SN}$ is the supernova rate.
This flux weakly depends on the 
interstellar medium number density $n_\mathrm{ISM}$ and the  
supernova explosion energy $E_\mathrm{SN}$. 
We recalculated the numerical factor in Eq. (53) using the $\Sigma -D$ relation for 
supernovae in the starburst galaxy M82 (Huang et al. \cite{huang94}). It is a factor 
of two larger than the factor 
obtained by Ulvestad (\cite{ulvestad82}). That means that supernovae in starburst 
galaxies are slightly brighter than in normal galaxies. 
This number is in a good agreement with radio observations of 
galaxies (cf. Condon \cite{condon92} for a review) and theoretical estimates 
(Lisenfeld \& V\"olk \cite{lisenfeld00}). 

We have performed the modeling of the starburst wind with a radius $R_g=300$ pc 
and have calculated synchrotron and thermal emission using the formulae given in 
this and previous sections. The choice of this radius was done in accordance with 
the size of the region where the radio emission of NGC253 is concentrated 
(Ulvestad \cite{ulvestad00}).

For the adiabatic index of the random magnetic field we use the adiabatic index 
of Alfv\'en waves $\gamma _m=3/2$. Since cosmic ray and Alfv\'en wave pressures 
are not negligible (see below) we also used a gas adiabatic index $\gamma =3/2$. 
The temperature of the absorbing gas was fixed at $8\cdot 10^3$ K. The infrared 
luminosity of the nucleus of NGC 253 was taken as $L_{\mathrm{IR}}=1.24\cdot 
10^{10}L_\odot $ (Radovich et al. \cite{radovich01}).  For the starlight 
luminosity, we used the luminosity of the bulge of NGC 253, 
$L_\mathrm{B}=0.39\cdot 10^{10}L_\odot $ (Pence \cite{pence80}). Also a source 
proton-to-electron ratio $K=70$ was chosen. Finally, a maximum energy $E_{\max 
}=10^6$ GeV of the cosmic ray protons was assumed. The results are shown in Fig.2 
and Fig.3.

The contour plot of the 1.4 GHz brightness distribution modeled is shown in 
Fig.2. The radio distribution was convolved with a Gaussian beam with FWHM 
$5.1''\times 3.8''$ ($62\times 46$ pc at 2.5 Mpc distance), and an inclination 
angle of $i=78^\circ $ was used. The distribution is slightly asymmetric relative 
to the horizontal axis due to free-free absorption. The comparison with the 
observed distribution (Mohan et al. \cite{mohan05}) permits us to estimate the 
wind velocity. We found that the wind velocity $u_\infty $ should be larger than 
500 km s$^{-1}$. For smaller velocities the brightness drops too fast with height 
above midplane because of strong synchrotron and inverse Compton losses. 
We have found a satisfactory fit for $L_{CR}=2.94\cdot 10^{41}$erg 
s$^{-1}$, $u_\infty =900$ km s$^{-1}$, $\tau _0=0.05|\cos{i}|$, $J_0=1.0$ Jy and 
$B_0=50$ $\mu $G. The magnetic field pressure was taken close to the cosmic ray 
pressure. The value of the radio flux $J_0$ from SNRs corresponds to the 
supernova rate $\nu _\mathrm{SN}=0.06$ yr$^{-1}$ which is intermediate between 
the value $\nu _\mathrm{SN}=0.03$ yr$^{-1}$ of Matila \& Meikle (\cite{matila01}) 
and $\nu _\mathrm{SN}=0.1$ yr$^{-1}$ (Ulvestad \& Antonucci \cite{ulvestad97}). 
Finally, a
mean number density $n_\mathrm{ISM}=100$ cm$^{-3}$ of the interstellar medium in 
NGC253 (Paglione et al. \cite{paglione95}), and a mean energy of supernova 
explosions $E_\mathrm{SN}=10^{51}$ erg were assumed in Eq. (53).  

The uncertainty of the velocity derived in this way is larger from the side 
of higher velocities. For example, for $u_\infty =500$ km s$^{-1}$ the contour 
lines near the upper and lower parts of the plot are shifted by about one contour 
level in the midplane direction. A similar shift in the opposite direction 
corresponds to $u_\infty =2000$ km s$^{-1}$. The derived value $u_{\infty}=900$ 
km s$^{-1}$ produces a good similarity between observed and modeled radio 
distribution in the southern part of the outflow. We think that the radio 
distribution in the northern part of the outflow should be explained by using a 
more complex model with slightly larger and variable wind velocities along the 
base (for example, another function $U(R_0)$ might be used). This can produce 
spur-like features in the observed radio distribution in the northern part of the 
outflow. Another possible explanation is limb-brightening due to the superwind's 
lateral interaction with halo gas which is not included in our model. 

The spectrum of the NGC 253 
nuclear region is shown in Fig.3 (solid line). The data points with 
the same resolution $33''\times 21''$ ($400\times 255$ pc at 2.5 Mpc distance) are   
taken from Carilli (\cite{carilli96}).  

The total radio flux of SNRs amounts to about one third of the total radio flux 
at 1.4 GHz. The thermal emission amounts to one third of the total intensity at 
10 GHz. This is in good agreement with results for the separation of the thermal 
and nonthermal emission for the whole galaxy NGC 253 (Niklas et al. 
\cite{niklas97}). The fit can be better if a smaller temperature of the emitting 
gas is used. It should be noted that Israel \& Mahoney (\cite{israel90}) found 
that the flattening of the radio spectra of galaxies at small frequencies can be 
explained by free-free absorption in a hypothetical cold ionized gas.

The observed synchrotron intensity mainly constrains the ratio of the cosmic 
ray power $L_{\mathrm{CR}}$ to the wind velocity. The wind velocity and cosmic 
ray power influence the spectral shape of the radio emission only weakly. 

The cosmic ray luminosity found corresponds to the cosmic ray pressure at the 
wind base (at $R=R_g$): 
\begin{equation}
p_{CR}=\frac {\gamma }{8(2\gamma -1)}\frac {L_{CR}}{\pi u_\infty R^2_g}
\end{equation}
This gives 
$p_{\mathrm{CR}}=1.08\cdot 10^{-10}$ erg cm$^{-3}$ at the wind base at $R_0=R_g$. 
The total nonthermal (cosmic ray + Alfv\'en waves) luminosity is about 
$5.6\cdot 10^{41}$ erg s$^{-1}$. 
\begin{figure}[t]
\includegraphics[width=5.5cm, angle=270]{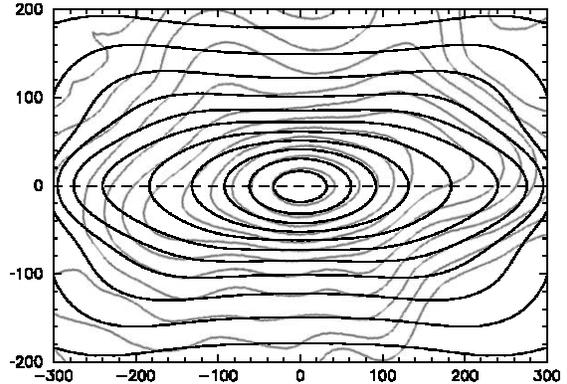}
\caption{Radio brightness distribution of NGC 253 at 1.4 GHz calculated in our 
model. 
The modeled radio intensity 
was convolved with a Gaussian beam of FWHM 
$5.1''\times 3.8''$ ($62\times 46$ pc at 2.5 Mpc distance). 
The {\it solid} 
contour levels are a geometric 
progression in 1.7, ending at 0.446 Jy/beam in the center. The projected distance 
in pc is shown on 
both axes. The observed brightness distribution at 1.4 GHz 
(Mohan et al. \cite{mohan05}) is shown by the
{\it gray scale} contours. 
}
\end{figure}

\begin{figure}[t]
\includegraphics[width=5.0cm, angle=270]{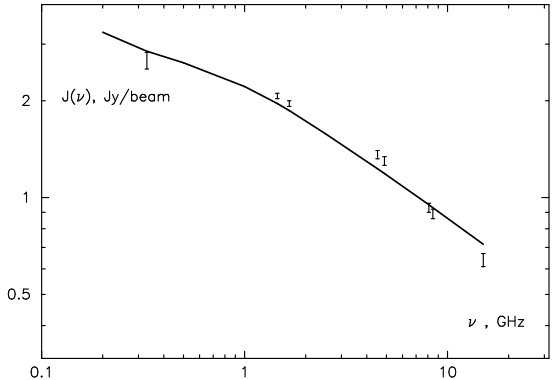}
\caption{Spectrum of the nuclear region of NGC 253 ({\it solid curve}), 
calculated in our model. The modeled radio intensity 
was convolved with a Gaussian beam of FWHM $33''\times 21''$ 
($400\times 255$ pc at 2.5 Mpc distance). The data points are taken from 
Carilli (\cite{carilli96}).}
\end{figure}

\section{Estimates of the outflow parameters}

We have modificated the Chevalier \& Clegg model (\cite{chevalier85}) for a thin 
starburst region (cf. Sect.3). In the particular case of the analytical solution 
(29),(30),(45),(46), the Eqs. (26) -- (28) can be rewritten as

\[
u_{\infty }=\frac \gamma {\gamma -1}u(h)=5.63\cdot 10^8 \mathrm{cm\ s}^{-1}
\]
\begin{equation}
\times \left( \frac {\dot{E}}{10^{43}\mathrm{ erg\ s}^{-1}}\right) ^{1/2}
\left( \frac {\dot{M}} {M_\odot \ \mathrm{yr}^{-1}}\right) ^{- 1/2}, 
\end{equation}  
\[
\rho (h)=1.88\cdot 10^{-27}\mathrm{g\ cm}^{-3}\frac {\gamma }{2\gamma -1} 
\left( \frac {\dot{E}}{10^{43}\mathrm{ erg\ s}^{-1}}\right) ^{-1/2}
\]
\begin{equation}
\times \left( \frac {\dot{M}}{M_\odot \ \mathrm{yr}^{-1}}\right) ^{3/2}
\left( \frac {R_g}{1\mathrm{\ kpc}}\right) ^{-2} 
\left( \frac {R_0}{R_g}\right) ^{-\frac {2\gamma }{2\gamma -1}}, 
\end{equation}   
\[
p(h)=2.98\cdot 10^{-10}\mathrm{erg\ cm}^{-3}\frac {\gamma -1}{\gamma ^2} 
\left( \frac {\dot{E}}{10^{43}\mathrm{ erg\ s}^{-1}}\right) ^{1/2}
\]
\begin{equation}
\times \left( \frac {\dot{M}}{M_\odot \ \mathrm{yr}^{-1}}\right) ^{1/2}
\left( \frac {R_g}{1\mathrm{\ kpc}}\right) ^{-2} 
\left( \frac {R_0}{R_g}\right) ^{-\frac {2\gamma }{2\gamma -1}}, 
\end{equation} 
Here $\dot{E}$ and $\dot{M}$ are the total energy and mass production rates of 
the starburst disk at $R<R_g$. 

The energy production rate is determined mainly by supernovae. The upper limit for 
the supernova rate in the nucleus of NGC 253 is 0.3 yr$^{-1}$ (Ulvestad \& Antonucci 
\cite{ulvestad97}). It seems that the actual number is not larger than 0.1 yr$^{-1}$ because 
of the absence of a new supernova during the last years. 
For a supernova rate 0.06 yr$^{-1}$ in NGC 253 the energy 
production rate $\dot{E}=1.9\cdot 10^{42}$ erg s$^{-1}$ under the assumption 
of negligible radiative losses of the gas. 
The nonthermal luminosity found in the previous section is 
about one quarter of this number.  

Since part of the supernova energy can be lost radiatively, the energy input may 
be smaller. The X-ray observations give an the energy input $\dot{E} \approx 
10^{42}$ erg s$^{-1}$ (Strickland et al. {\cite{strickland02}}) and an average 
wind velocity $u$ between 400 and 500 km s$^{-1}$ that is between the velocity 
300 km s$^{-1}$ at the wind base and the asymptotic velocity 900 km s$^{-1}$ 
found in the previous section.

For this case we found the mass-loss rate $\dot{M}=3.9M_\odot$ yr$^{-1}$, and the 
total pressure $p=4.6\cdot 10^{-10}$ erg cm$^{-3}$ at $R=R_g$ and a factor of 15 
larger at a distance of 50 pc from the nucleus. These numbers should be compared 
with the cosmic ray and magnetic pressures derived in the previous section 
$p_{\mathrm{CR}}=1.1\cdot 10^{-10}$ erg cm$^{-3}$ and $B^2_0/8\pi =1.0\cdot 
10^{-10}$ erg cm$^{-3}$ respectively. Therefore the nonthermal pressure is almost 
one half of the total pressure. The rest can be attributed to the thermal gas 
pressure, that is $p_g=2.5\cdot 10^{-10}$ erg cm$^{-3}$. The corresponding mass 
density is $\rho _g=3.8\cdot 10^{-25}$ g cm$^{-3}$. The corresponding temperature 
of a one-phase fully ionized gas is about $5 \cdot 10^{6}$ K.

This is not surprising. If 15 procent of the supernova energy is 
converted into cosmic ray energy and the rest is partially radiated, then the gas pressure 
may be comparable with the cosmic ray pressure. In this case it is reasonable to use 
a smaller adiabatic index $\gamma <5/3$ of the gas to include cosmic rays and 
magnetic fields.  

In reality, a one-phase thermal gas with a temperature of several million degrees 
is excluded by the X-ray observations, since it would produce too large an amount 
of X-rays (Strickland et al. {\cite{strickland02}}). The main fraction of the 
thermal energy may rather be contained in an even hotter gas component, and the 
main part of the mass may then be contained in gas with a temperature below one 
million degrees.

We can even consider an extreme case by assuming that the gas radiates all its 
energy. The superwind flow is then driven by cosmic rays and Alfv\'en waves (or 
turbulent magnetic fields) alone. Using a nonthermal luminosity $\dot{E}=5.6\cdot 
10^{41}$ erg s$^{-1}$, as found in the last section, we then obtain a mass-loss 
rate of $\dot{M}=2.2\ M_{\odot }$ yr$^{-1}$.

\section{Conclusion}

We have presented a simple analytical model for a steady state axially symmetric 
gas-dynamic outflow from a starburst galaxy. In the case of a power-law radial 
distribution of the gas pressure at the wind base self-similar solutions can be 
found. For the special choice of the power-law index $\delta =-2\gamma /(2\gamma 
-1) \approx 1.5$ we found an explicit analytical solution. This value of the 
index seems reasonable, since many starbursts show a strong concentration of the 
energy release towards the center.

The solution found has less steep pressure and density profiles in comparison 
with the Chevalier \& Clegg model (Chevalier \& Clegg \cite{chevalier85}). It is 
well known that the Chevalier \& Clegg solution is in agreement with the X-ray 
brightness of M82 at distances larger than 1-2 kpc from the nucleus, whereas at 
smaller distances the observed brightness relief is lower (e.g. Fabbiano 
\cite{fabbiano88}). This may be understood in the framework of our model. At 
distances smaller than the starburst region gas collimation is possible. This 
results in weaker dependencies on radius. At larger distances the wind geometry 
is roughly spherically symmetric, perpendicular forces play no role and the 
Chevalier \& Clegg model is more adequate.

In reality the power-law pressure dependence can not be 
valid at small radii either,    
i.e. there exists some minimum radius. The wind originating inside this radius is 
not described by the self-similar solution. However, for a small minimum radius 
it blows in a narrow region near the axis. Therefore this restriction appears not 
to be essential.

The solution found was applied to the modeling of the radio distribution of the 
central part ($R<300$ pc) of the starburst galaxy NGC 253. This galaxy contains a 
very bright radio nucleus with a flat spectrum (with spectral index rnaging from 
$-0.43$ to $-0.17$ at different frequencies) and a less bright extended radio 
disk with a steeper spectrum of spectral index -0.7 (Hummel et al. 
\cite{hummel84}, Carilli et al. \cite{carilli92}).

The propagation of the radio electrons can be purely advective in the central 
part of NGC 253 since the gas velocity is rather high (about 1000 km s$^{-1}$). 
It appears that this velocity is smaller in the outer part of the galactic disk 
($R>1$ kpc) where the spectrum is steeper. It is possible that the outer part 
contains a cosmic ray driven galactic wind which has small velocity at the wind 
base (cf. Breitschwerdt et. al. \cite{breitschwerdt91}; Zirakashvili et al. 
\cite{zirakashvili96}) and where the diffusion of particles results in the 
spectral steepening.

Our model is applicable for the central superwind region ($R<300$ pc) of NGC253 
which contains the central starburst. NGC253 has also a comparable energy release 
at larger distances. However this is distributed over an area which is at least a 
factor 100-1000 larger than the central starburst. Our model does not describe 
that region which contains a quasi-static galactic halo or a slow cosmic-ray 
driven galactic wind. It also does not describe the interaction of the gas in 
this region with the superwind flow.

We have obtained a satisfactory fit for the radio brightness distribution and for 
the radio spectrum of the nucleus of NGC 253. This permits us to estimate the 
parameters of the outflow: the wind velocity and mass loss rate of the starburst.

From a physics point of view these quantities should be determined from a 
theoretical model which describes how thermalised supernova ejecta and the 
interstellar gas heated by supernova shocks radiate some part of the energy and 
form an outflow from the starburst disk. In the absence of such a theory we limit 
ourself to the determination of these quantities from observations.

 
We found an asymptotic velocity of 900 km s$^{-1}$ from this inner region.  For 
an energy input of $10^{42}$ erg s$^{-1}$ in the starburst nucleus of NGC253 we 
obtain a mass loss rate of $4 M_{\odot}$ yr$^{-1}$. This number drops down to $2 
M_{\odot}$ yr$^{-1}$ for the extreme case when the gas radiates almost all its 
energy. Even this case does not contradict the X-ray observations, since it appears 
that the X-rays are produced by a hot gas that is formed from the interaction 
between the superwind and the circumstellar medium (halo gas, clouds etc.) (cf. 
Strickland et al. \cite{strickland02}).

The cosmic ray production power is estimated as 15 procent of the supernova 
power. We found that cosmic ray and magnetic pressures are comparable with the 
gas pressure. Therefore we can conclude that the starburst outflow in NGC 253 is 
collectively driven by hot gas, cosmic rays and turbulent magnetic fields.
 
%

\begin{acknowledgements}
 
We thank the anonymous referee for a number of valuable suggestions regarding the 
interpretation of the theoretical results.
\end{acknowledgements}
%
%

\end{document}